%% file: main.tex
\documentclass[]{spie}  

\usepackage{amsmath,amsfonts,amssymb}
\usepackage{graphicx}
\usepackage[colorlinks=true, allcolors=blue]{hyperref}
\usepackage{setspace}
\usepackage{xcolor}
\usepackage{titlesec}
\usepackage{graphicx}
\usepackage{float}
\usepackage{subcaption}
\usepackage{booktabs}
\usepackage{enumitem}
\usepackage{comment}
\usepackage{amsmath}

\title{Comparative Study of Hollow-Core and Standard Optical Fibers for Astronomy}

\author[a]{Malak Galal}
\author[a]{Oliver Pineda Suárez}
\author[b,c]{Frédéric Gérôme}
\author[b,c]{Benoit Debord}
\author[b,c]{Fetah Benabid}
\author[a]{Jean-Paul Kneib}
\affil[a]{Institute of Physics, Laboratory of Astrophysics, Ecole Polytechnique Federale de Lausanne (EPFL), Observatoire de Sauverny, CH-1290 Versoix, Switzerland}
\affil[b]{1GPPMM Group, XLIM Institute, CNRS UMR 7252,University of Limoges, Limoges, France}
\affil[c]{GLOphotonics, 123 Avenue Albert Thomas, Limoges,
France}

\authorinfo{Further author information: (Send correspondence to M. G.)\\M. G.: E-mail: malak.galal@epfl.ch, Telephone: +41 21 693 92 91}

\begin{document} 
\maketitle

\begin{abstract}
Efficient light transmission in the blue-visible regime remains a major limitation for fiber-fed astronomical spectrographs, where low photon flux and the intrinsic attenuation of conventional silica fibers reduce survey sensitivity and depth. Inhibited-coupling hollow-core fibers (IC-HCFs) with reduced surface roughness offer a promising alternative, providing guidance predominantly in air and enabling significantly lower loss across the visible spectrum. In this study, we present a comparative evaluation of IC-HCFs against standard multi-mode fibers used in current astronomical instrumentation. We assess the throughput loss that occurs due to bending, twisting, or pinching of the optical fibers when moved using one of the robotic fiber-positioner prototypes designed for next-generation telescopes. These measurements quantify the performance gains offered by IC-HCFs for blue-sensitive spectroscopy and assess their suitability for integration into future survey facilities.
\end{abstract}


\input{a_Sec1}
\input{a_Sec2}

\input{a_Sec3}
\input{a_Sec4}
\input{a_Sec5}

\input{a_Sec6}

\acknowledgments 
 
The authors would like to acknowledge MPS MicroPrecision Systems AG for the development of the robotic positioner module, as well as thank Dr. Claire Poppett and Dr. Gregory Jasion for their insights and the fruitful discussions conducted, and Alexandre Gorse and Dr. Frédéric Delahaye from GLOphotonics for their help in the hollow-core fiber assembly.  

\bibliography{report} 
\bibliographystyle{spiebib} 

\end{document}

%% file: a_Sec1.tex
\section{INTRODUCTION}
\label{sec:intro}  

Wide-field fiber-fed spectroscopic surveys have transformed observational cosmology by enabling the simultaneous acquisition of spectra for thousands of astronomical sources. From the pioneering Sloan Digital Sky Survey to current facilities such as Dark Energy Spectroscopic Instrument, Euclid, and 4MOST, multiplexed optical fiber systems combined with robotic positioners have become the cornerstone of large-scale structure studies and precision cosmology. These instruments efficiently transport light from the telescope focal plane to highly stable spectrographs, allowing the construction of three-dimensional maps of the Universe and the measurement of galaxy redshifts on unprecedented scales. The next generation of Stage-5 spectroscopic facilities, including the Chinese MUST \cite{zhao2024multiplexed}, the American Spec-S5 \cite{besuner2025spectroscopic}, and the European WST \cite{mainieri2024wide} projects, aims to increase multiplexing by an order of magnitude and to probe the redshift range $2<z<5$, corresponding to the epoch of peak galaxy assembly. Achieving these ambitious science goals requires a substantial improvement in end-to-end throughput, particularly at blue visible wavelengths, where key spectral features of high-redshift galaxies are observed.

In current instruments, light transport is almost exclusively performed using broadband multi-mode silica fibers. Although these fibers provide excellent performance in the red and near-infrared, their transmission efficiency rapidly decreases toward shorter wavelengths due to intrinsic material absorption and Rayleigh scattering. For the tens of meters of fiber required in modern telescope layouts, this wavelength-dependent attenuation leads to significant photon losses and directly impacts survey speed, sensitivity, and the achievable signal-to-noise ratio for faint sources. As a result, the fiber link has become one of the dominant throughput limitations in blue-sensitive astronomical spectroscopy.

Inhibited-Coupling (IC) or Anti-Resonant (AR) Hollow-Core Fibers (HCFs) offer a fundamentally different light-guiding mechanism that can address this bottleneck \cite{melli2026disentangling}. By confining light predominantly in an air-filled core rather than in solid glass, these fibers strongly reduce the interaction between the optical field and the material, enabling ultra-low attenuation, broad transmission windows, and low nonlinear and dispersive effects. In the field of telecommunications, AR-HCFs have already demonstrated losses below those of conventional silica fibers in specific wavelength ranges, while recent designs have shown remarkable performance extending into the visible regime. Their micro-structured geometry also provides a powerful design parameter space in which the transmission band, modal content, and effective numerical aperture can be tailored to match the requirements of astronomical instruments.

Despite this rapid technological maturity and their growing impact in other areas of photonics \cite{gao202540, mulvad2022kilowatt, galal2023study, taranta2020exceptional, liu2018nonlinearity}, hollow-core fibers have so far seen very limited adoption in astronomy. Early studies based on photonic bandgap designs were hindered by high loss, narrow bandwidth, and fabrication complexity, preventing practical implementation. The emergence of anti-resonant guidance has fundamentally changed this landscape, opening the possibility of a new class of low-loss, high-fidelity fiber links for astronomical applications. However, a comprehensive experimental and system-level evaluation of IC-HCFs/AR-HCFs in the context of fiber-fed spectrographs, addressing key parameters such as throughput, focal ratio degradation, numerical aperture matching, and integration within robotic positioners, has not yet been carried out. The only results that have been reported so far in relation to next-generation robotic positioners are purely mechanical studies to evaluate the performance of positioners themselves \cite{galal2025prototyping}.

In this paper, we investigate the feasibility of using inhibited-coupling hollow-core fibers as a next-generation light-transport solution for astronomical instrumentation. We assess their potential to overcome the blue-wavelength transmission limitations of conventional multi-mode fibers and to enhance the performance of future high-multiplex spectroscopic facilities. We do this by integrating them on state-of-the-art 6.2-mm pitch fiber positioner modules to assess how the movement of the robotic positioners impacts the optical performance of such fibers. By bridging recent advances in hollow-core fiber technology with the stringent requirements of modern telescope systems, this work aims to establish whether IC-HCFs can enable faster surveys, improved sensitivity to faint high-redshift galaxies, and a more efficient realization of Stage-5 cosmological experiments.

%% file: a_Sec2.tex
\section{ONGOING MAJOR FIBER-FED ASTRONOMICAL SURVEYS}

\begin{figure}[!b]
    \centering
    \begin{subfigure}[b]{0.30\textwidth}
        \centering
        \includegraphics[width=\textwidth]{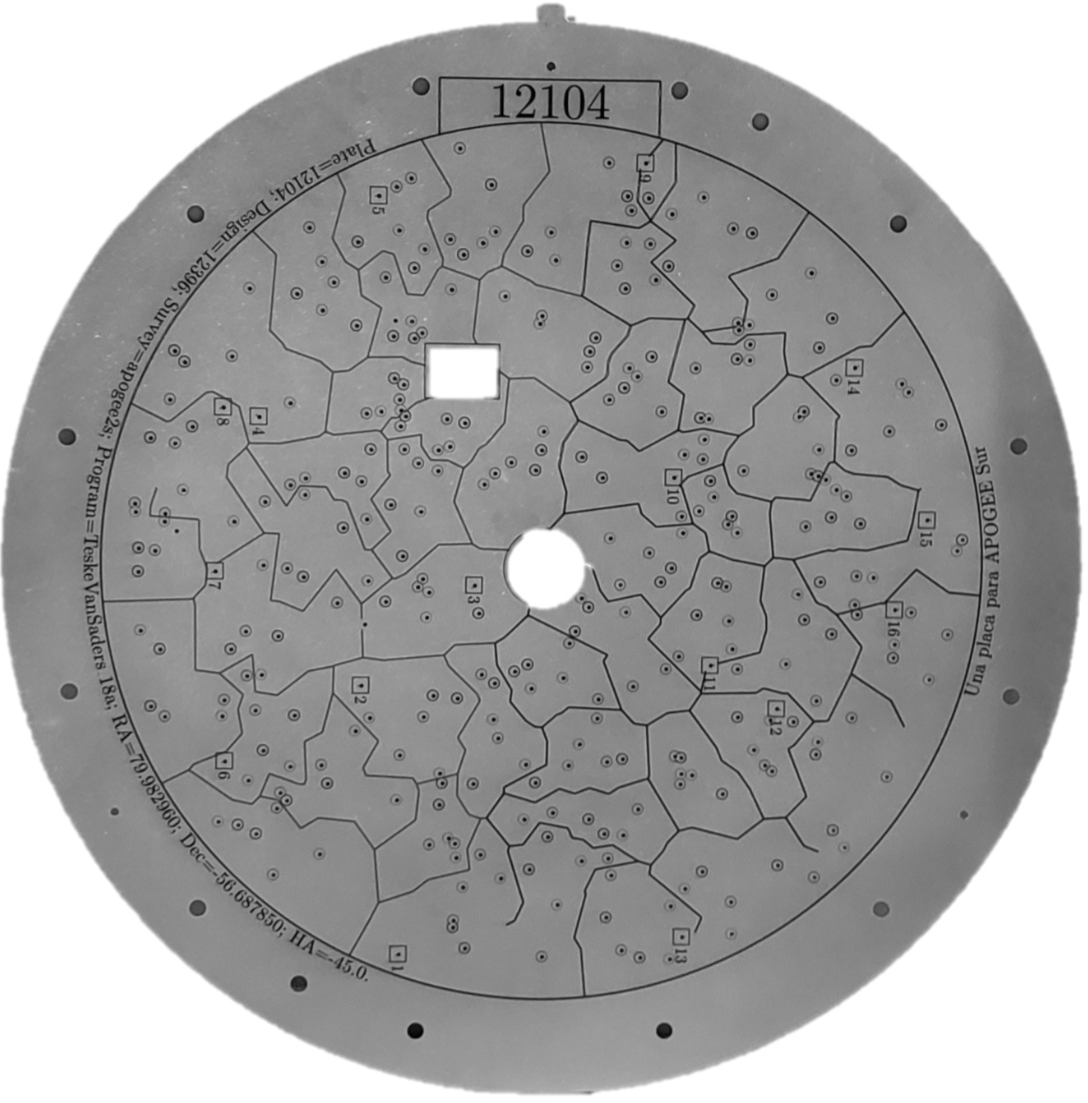}
        \caption{}
        \label{fig:plug_plate_focal_plane}
    \end{subfigure}
    \hspace{0.02\textwidth} 
    \begin{subfigure}[b]{0.35\textwidth}
        \centering
        \includegraphics[width=\textwidth]{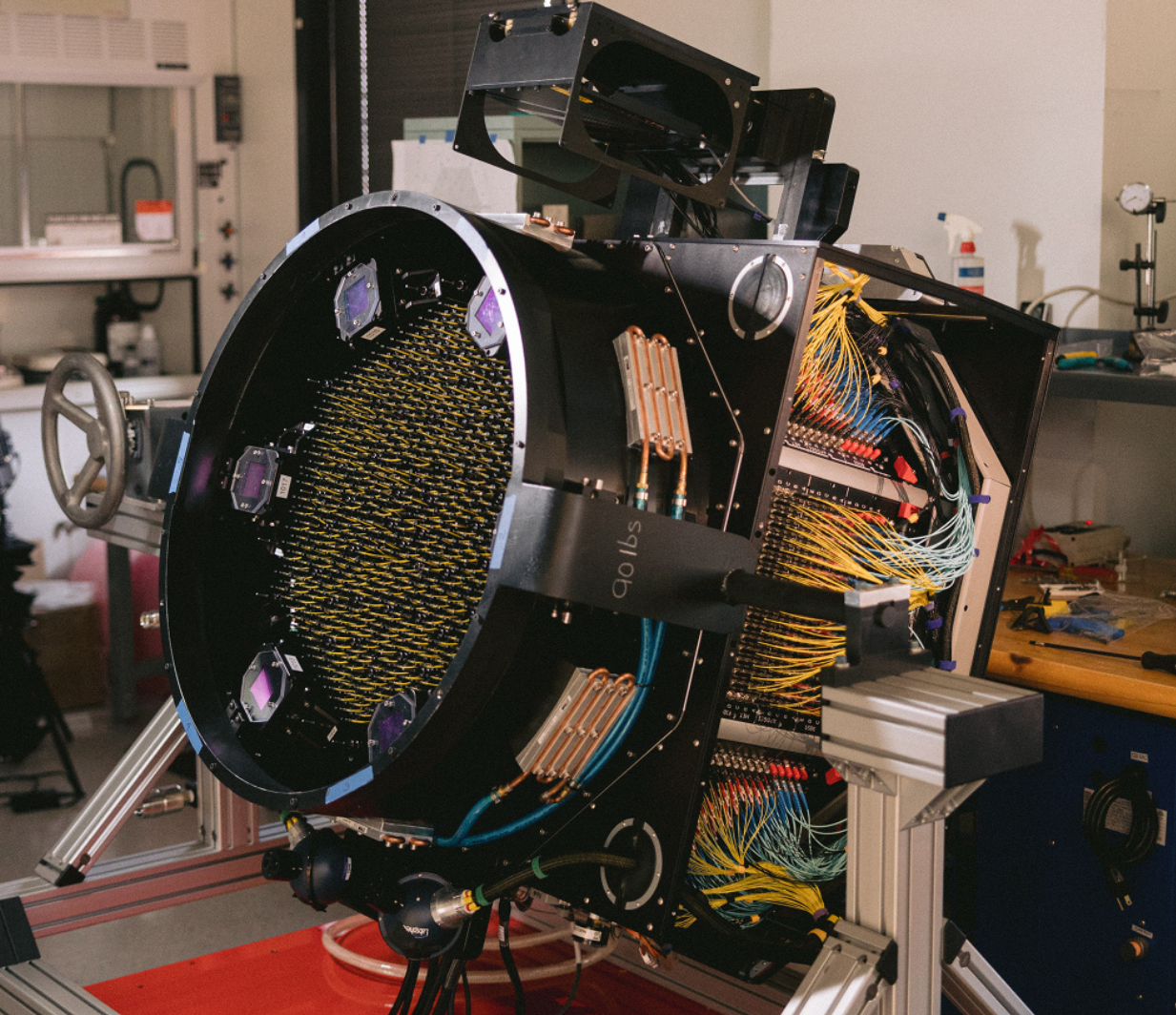}
        \caption{}
        \label{fig:robotic_focal_plate}
    \end{subfigure}
    \caption{Focal plates for optical fibers of SDSS surveys; (a) Manual hand-plug focal plate \cite{sdss_apogee_observing}; (b) Robotic focal plate \cite{sdss_fps}.}
    \label{fig:focal_plate}
\end{figure}

\begin{figure}[!t]
    \centering
    \begin{subfigure}[b]{0.48\textwidth}
        \centering
        \includegraphics[width=\textwidth]{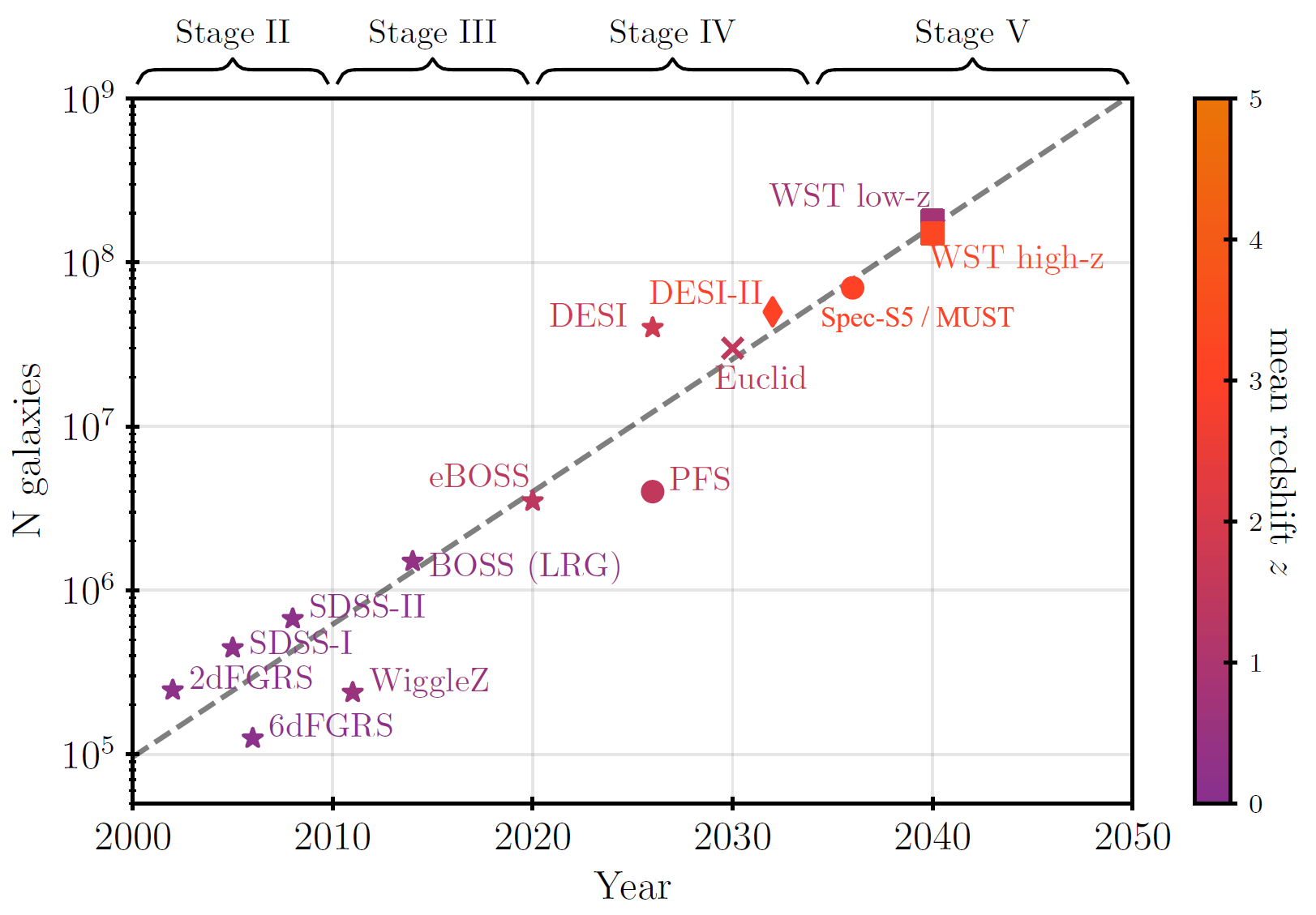}
        \caption{}
        \label{fig:time_vs_galaxies_surveyed}
    \end{subfigure}
    \hspace{0.015\textwidth} 
    \begin{subfigure}[b]{0.38\textwidth}
        \centering
        \includegraphics[width=\textwidth]{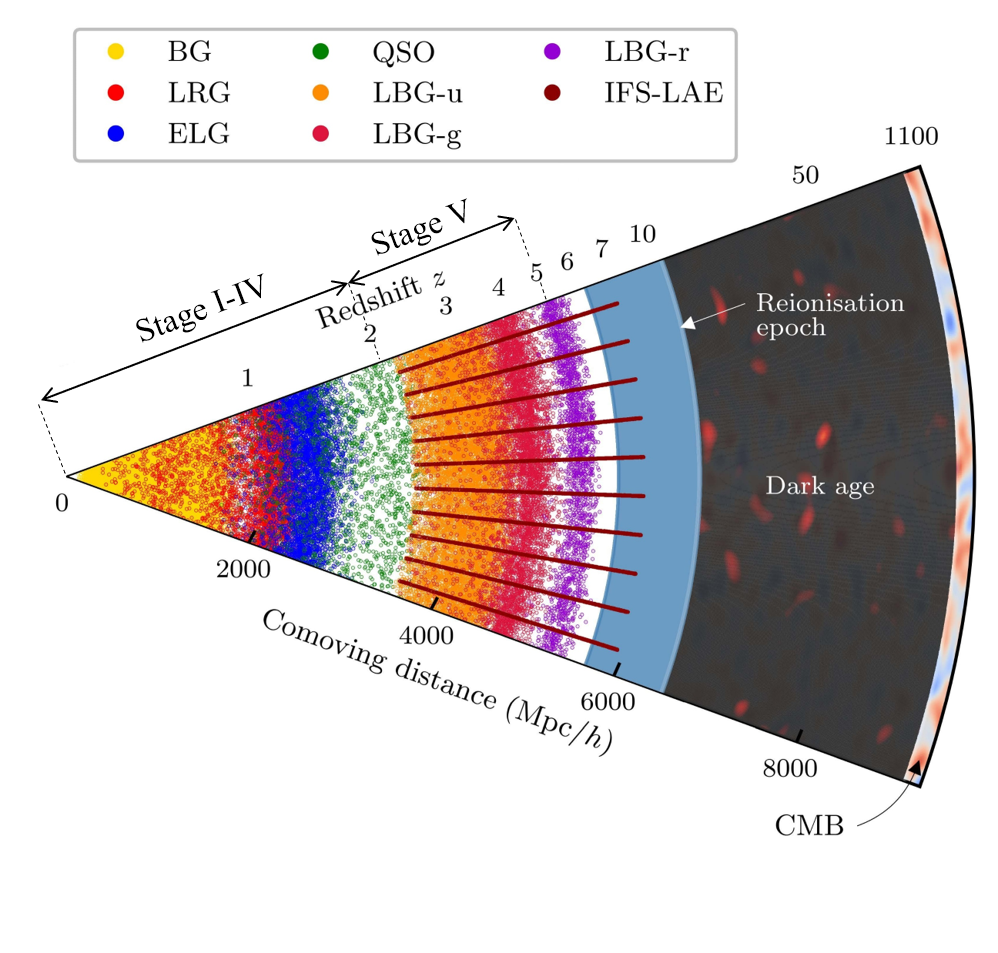}
        \caption{}
        \label{fig:redshifts_schematics}
    \end{subfigure}
    \caption{Graphs highlighting the interest of the astronomical community to pursue building Stage-5 instruments to enhance our comprehension of the Universe; (a) Progress in wide cosmology spectroscopic redshift surveys of the year 2000 to 2050 \cite{mainieri2024wide}; (b) Schematic using the light cone representation to show the range of future Stage-5 MOS instruments \cite{mainieri2024wide}.}
    \label{fig:stageV_surveys}
\end{figure}

Over the past few decades, fiber-fed spectroscopic instruments have revolutionized observational astronomy by enabling simultaneous light collection from thousands of celestial objects across wide fields of view. Modern systems rely on robotic fiber positioners, which have replaced the static plug-plate systems used in earlier surveys as seen in Figure \ref{fig:focal_plate}. In traditional designs, optical fibers had to be manually plugged into pre-drilled metal plates corresponding to specific sky coordinates, which is a time-consuming process that could take several hours and required physically mounting and unmounting plates between observations. In contrast, robotic positioners can automatically and precisely place each fiber at its designated target location within minutes, dramatically increasing observing efficiency and survey flexibility.
Optical fibers transport light from the telescope focal plane to highly stable, temperature-controlled spectrographs, allowing precise measurements of galaxy spectra, redshifts, and chemical compositions. 
This architecture has been fundamental to large-scale spectroscopic surveys, enabling the construction of three-dimensional maps of the Universe and transforming our ability to probe its structure and evolution. The combination of wide-field telescopes, robotic positioners, and multiplexed fiber-fed spectrographs has thus become the cornerstone of modern cosmology and extragalactic astronomy.
The Sloan Digital Sky Survey (SDSS) \cite{york2000sloan} marked a transformative milestone in observational cosmology through its 20-year effort of massive spectroscopic mapping using the 2.5-meter Sloan Telescope. This pioneering project produced the first comprehensive three-dimensional map of the Universe, inspiring a new generation of large-scale spectroscopic surveys aimed at achieving even greater precision and coverage. The most recent phase, SDSS-V, extends this legacy by operating from two telescopes, one in the northern hemisphere and one in the southern hemisphere, each equipped with 500 optical fibers mounted on 500 individual robotic positioners. This fully automated configuration enables rapid field reconfiguration and continuous sky coverage across both hemispheres. Mapping the large-scale distribution of galaxies provides a fundamental tool for tracing the matter content of the Universe, studying its expansion history, and constraining the parameters of the cosmological model and fundamental physics \cite{percival2007measuring}.
Building on this legacy, the Dark Energy Spectroscopic Instrument (DESI) \cite{levi2013desi, aghamousa2016desi} began operations in 2019 with 5000 robotic positioners and optical fibers and has been systematically surveying the sky over the past four years. Preliminary results from its first year of data suggest a potential deviation from the standard Lambda Cold Dark Matter ($\Lambda$CDM) model \cite{lodha2025desi}. 
In 2026, DESI will present scientific results based on an extensive dataset containing approximately 40 million galaxy and quasar redshifts (for z $<$ 1.6), providing unprecedented precision in cosmological measurements.
Two complementary projects—the ESA Euclid mission (launched in July 2023) \cite{mellier2024euclid} and the ESO 4MOST survey 
(first light obtained on October 21, 2025) \cite{de20124most}, will extend DESI’s reach both in redshift and sky coverage. Euclid aims to obtain over 50 million galaxy redshifts out to z $\sim$ 2, while 4MOST will collect an additional 8 million redshifts across the Southern hemisphere. Together, these ambitious programs will deliver a remarkably detailed view of the large-scale structure of the Universe. However, despite the combined progress of SDSS, DESI, Euclid, and 4MOST, a significant observational gap remains. The redshift range between 2 $<$ z $<$ 5, corresponding to the epoch of galaxy assembly and early structure formation, remains largely unexplored by these surveys. Addressing this gap will be essential to complete the three-dimensional mapping of the cosmos and to further refine our understanding of the Universe’s evolution (see Figure \ref{fig:stageV_surveys}).

\begin{figure}[H]
    \centering
    \includegraphics[width=0.7\linewidth]{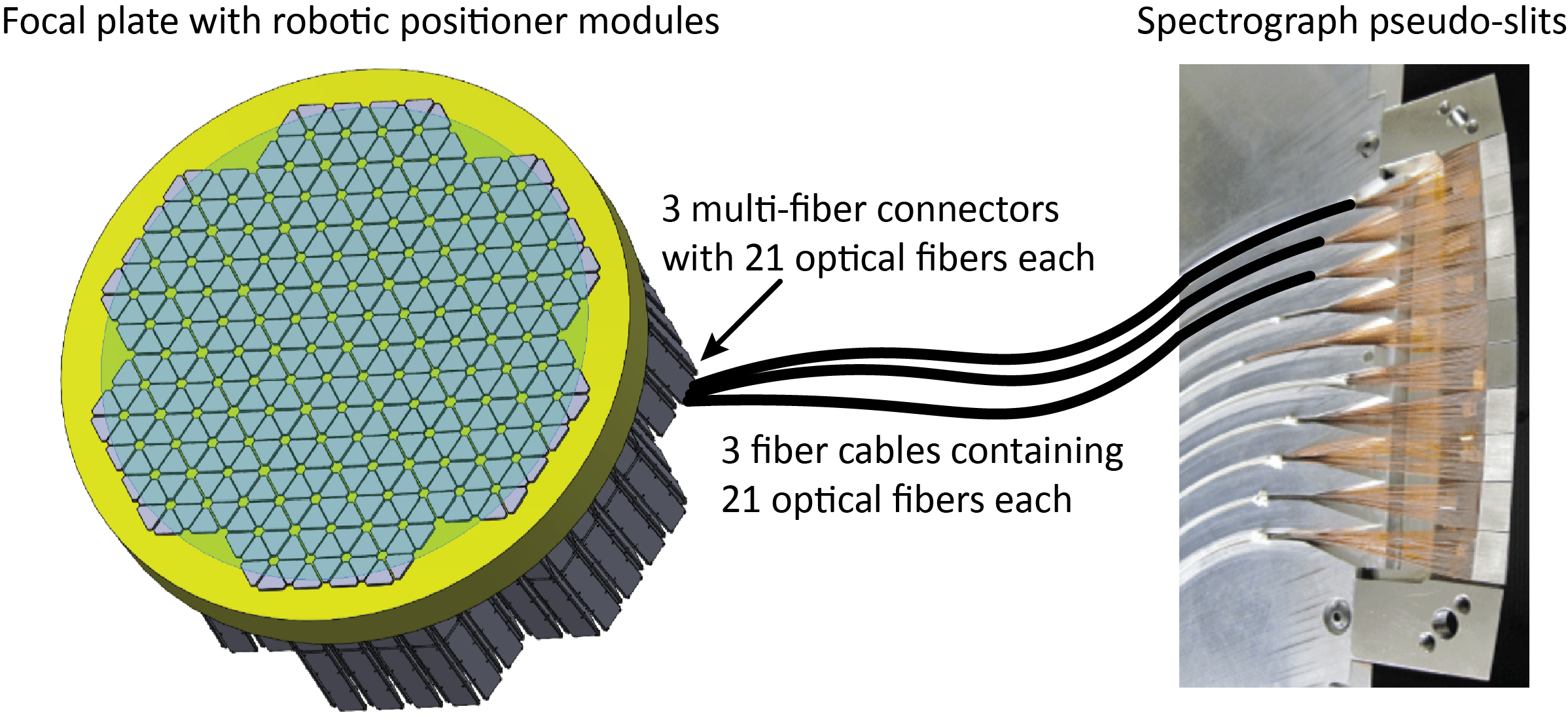}
    \caption{Schematic showing the focal plate with modules each having 63 robotic positioners and each positioner holding one optical fiber to be routed and be placed onto V-groves of the spectrograph pseudo-slits.}
    \label{fig:platewithmodules}
\end{figure}

A new generation of instruments, known as Stage-5 projects, is being developed to advance spectroscopic surveys toward having over 20,000 robotic positioners and optical fibers. These systems adopt a modular, maintenance-efficient architecture in which the focal plate is assembled from modules containing 63 positioners each, rather than mounting thousands of individual units \cite{SilberModule, rombach2024investigations}. A schematic of how the system would look like can be seen in Figure \ref{fig:platewithmodules}.

Each module can be pre-assembled, tested, and replaced as a single block, greatly simplifying integration, maintenance, and scalability. This design reduces downtime, improves reliability, and streamlines large-scale operations. When combined with innovations such as low-loss anti-resonant hollow-core fibers, Stage-5 telescopes will enable faster, more efficient, and higher-throughput spectroscopic mapping of the distant Universe.

%% file: a_Sec3.tex
\section{CURRENTLY EMPLOYED OPTICAL FIBERS IN ASTRONOMY}
\begin{figure}[!b]
    \centering
    \includegraphics[width=0.6\linewidth]{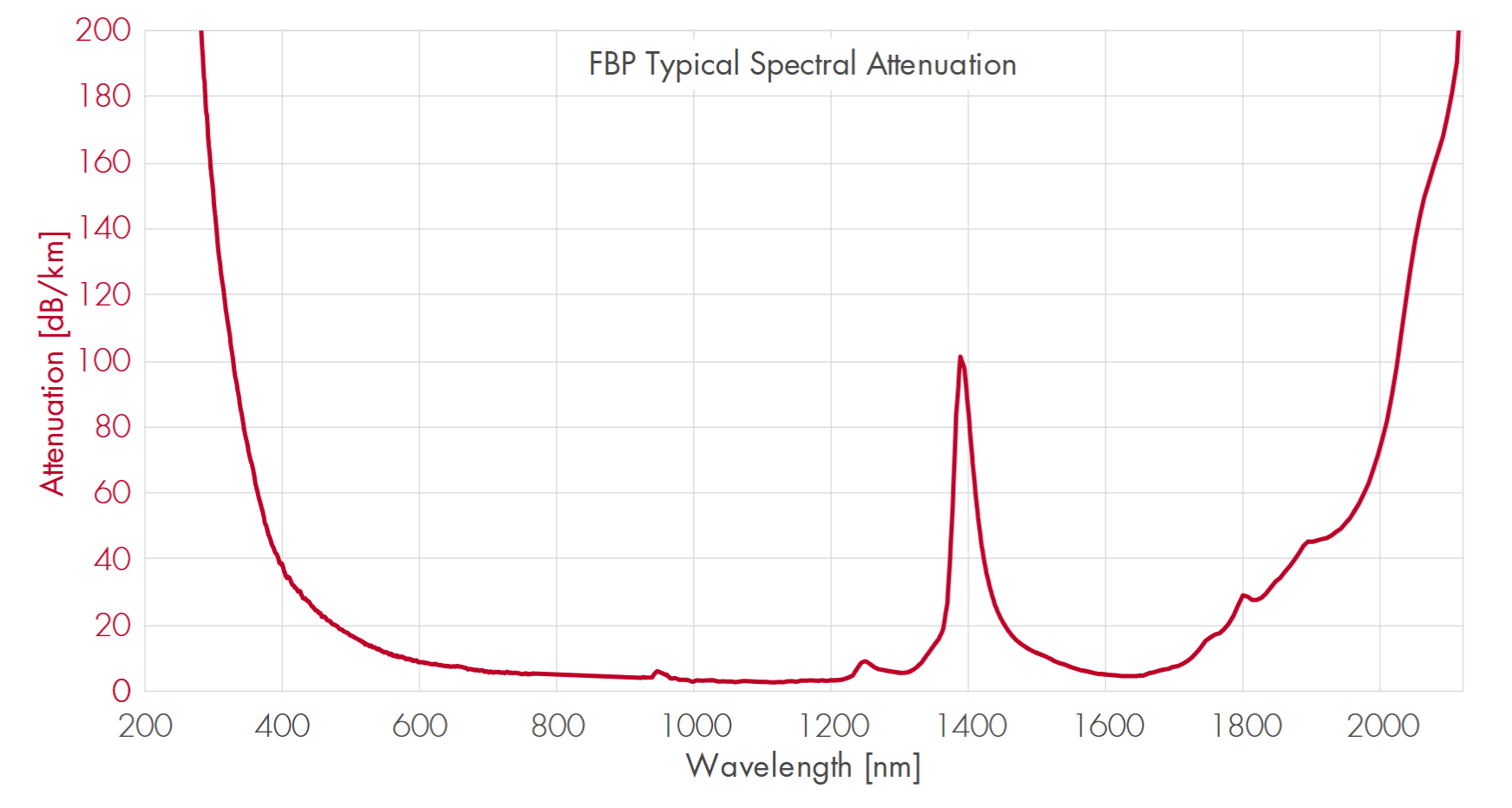}
    \caption{Graph showing wavelength spectrum of the FBP broad-band multi-mode optical  \cite{lasercomponents_fbp}.}
    \label{fig:FBP_polymicro_spectrum}
\end{figure}

Current astronomical instrumentation predominantly relies on conventional silica-based multi-mode optical fibers (with numerical aperture of 0.22) to transport light from telescope focal planes to spectrographs. Leading manufacturers such as Polymicro Technologies and CeramOptec have developed broadband fibers optimized for astronomical applications, among which the Polymicro FBP series is one of the most widely adopted. These fibers offer excellent throughput across the visible and near-infrared ranges, with low focal ratio degradation, good mechanical robustness, and long-term environmental stability, which are qualities that have made them the standard choice for major astronomical projects. 
Despite these advantages, silica fibers exhibit significant transmission losses in the blue part of the spectrum ($\lambda$ $<$ 450 nm) due to intrinsic absorption and Rayleigh scattering in the glass matrix. Even high-performance broadband fibers such as the FBP series show attenuation levels of around 1$\%$ per meter at short wavelengths. The full spectrum of commonly employed Polymicro FBP optical fibers in astronomy is shown in Figure \ref{fig:FBP_polymicro_spectrum}. 

The fiber system throughput efficiency for Polymicro FBP broad-band optical fibers have been reported by the DESI project \cite{poppett2024overview} and are shown in the table below:

\begin{table}[h!]
    \centering
    \caption{Fiber System throughput efficiency over 50 m of fiber length.}
    \label{tab:fiber_throughput}
    \begin{tabular}{l *{9}{c}} 
        \toprule
        \toprule
        & \multicolumn{9}{c}{Wavelength (nm)} \\
        & 360 & 375 & 400 & 500 & 600 & 700 & 800 & 900 & 980 \\
        \midrule
        Fiber System Throughput (\%) & 55.5 & 65.5 & 70.5 & 84.8 & 89.1 & 90.3 & 91.4 & 92.9 & 93.7 \\
        Throughput Loss (dB/km) & 51.14 & 36.76 & 30.34 & 14.22 & 9.98 & 9.07 & 8.00 & 6.42 & 5.64 \\
        \bottomrule
        \bottomrule
    \end{tabular}
\end{table}

For large fiber-fed instruments requiring tens of meters of fiber, these cumulative losses dramatically reduce the signal-to-noise ratio, especially for faint celestial targets. This wavelength-dependent limitation poses a major obstacle to deep spectroscopic surveys and studies of high-redshift galaxies, which rely on capturing blue-shifted emission and absorption lines.

Scientists and engineers in the astronomical community were eager to come up with different structures and forms of optical fibers to tackle the aforementioned issues. For example, hexabundles of optical fibers have been proposed for low-light astronomical applications \cite{bland2011hexabundles, bryant2014focal}. Such fibers provide spatially resolved spectra and increased observing efficiency by acting as mini integral-field units. However, they suffer from focal ratio degradation, inter-fiber cross-talk, and significant transmission losses, especially in the blue—limiting throughput for faint or blue-sensitive sources.

Other types of optical fibers explored by the community are multi-core fibers \cite{jovanovic2020astronomical}. These fibers offer the ability to collect light from multiple cores within a single cladding, enabling compact multi-object or integral-field spectroscopy with improved spatial sampling. However, they suffer from inter-core cross-talk, modal noise, and still experience significant transmission losses in the blue spectral range, limiting their overall efficiency for faint-object observations.

The concept of using hollow-core fibers for astronomy was explored more than a decade ago at Durham University. In the doctoral thesis by Claire Poppett completed in 2011 \cite{poppett2011new}, the use of photonic bandgap hollow-core fibers was studied as a potential replacement for silica fibers in astronomical light transport systems. Although these fibers successfully demonstrated light guidance predominantly through air, they suffered from intrinsic problems such as surface modes, limited operational bandwidth, and fabrication sensitivity. These issues led to losses that remained higher than those achievable in silica fibers, preventing their adoption in astronomical instruments. Consequently, the topic was largely abandoned in subsequent years, with little follow-up research within the astronomical instrumentation community \cite{dewitt2023transmission, 10.1117/12.3020535}. 

%% file: a_Sec4.tex
\section{HOLLOW-CORE FIBERS: STATE-OF-THE-ART}
In recent years, however, a new class of fibers known as anti-resonant (AR) or inhibited coupling (IC) hollow-core fibers has fundamentally altered the landscape. A review article published in 2023 \cite{jovanovic20232023} presented a roadmap for astrophotonics highlighting the potential of hollow-core fibers with more emphasis on AR-HCFs/IC-HCFs to be used in astronomy. Unlike photonic bandgap fibers, IC-HCFs employ an inhibited coupling guiding mechanism that confines light within an air-filled core over a broad wavelength range, while minimizing coupling to lossy surface modes \cite{roberts2005ultimate, poletti2014nested}. Their microstructured design, typically composed of a central hollow core surrounded by thin-walled capillaries, enables ultra-low loss, wideband transmission, and reduced modal distortion. 
Crucially, IC-HCFs/AR-HCFs have achieved record-breaking performance in telecommunications, surpassing conventional silica fibers in some wavelength ranges. Francesco Poletti and colleagues at the Optoelectronics Research Centre, University of Southampton, have been central to these developments. Their research has demonstrated how careful design of cladding geometry and wall thickness can suppress confinement and surface scattering losses to unprecedented levels. A seminal paper by Poletti et al. \cite{poletti2014nested} outlined the theoretical foundations for achieving sub-dB/km attenuation through anti-resonant designs. Follow-up experimental realizations achieved losses as low as 0.28 dB/km at 1550 nm \cite{jasion2020hollow} and later reaching 0.174 dB/km \cite{jasion20220} and in 2025 the loss reached is even below 0.1 dB/km in optimized nested-tube structures \cite{petrovich2025broadband}.

As mentioned above, even high-performing instruments such as DESI suffer from faint light transmissions, particularly in the blue regime. To date, no studies have evaluated the integration of IC-HCFs/AR-HCFs in astronomical systems to address this weak-signal transmission problem.

Given these limitations and the potential of IC-HCFs/AR-HCFs for exceptionally low attenuation, tunable spectral windows, and compatibility with modern fiber fabrication techniques, IC-HCFs/AR-HCFs represent a promising next-generation technology for high-throughput, low-loss fiber links in astronomy. Their adoption could significantly enhance the efficiency of future spectroscopic instruments, enabling deeper surveys, improved sensitivity, and higher spectral fidelity across the entire visible range.

\begin{figure}[!t]
    \centering
    \begin{subfigure}[b]{0.50\textwidth}
        \centering
        \includegraphics[width=\textwidth]{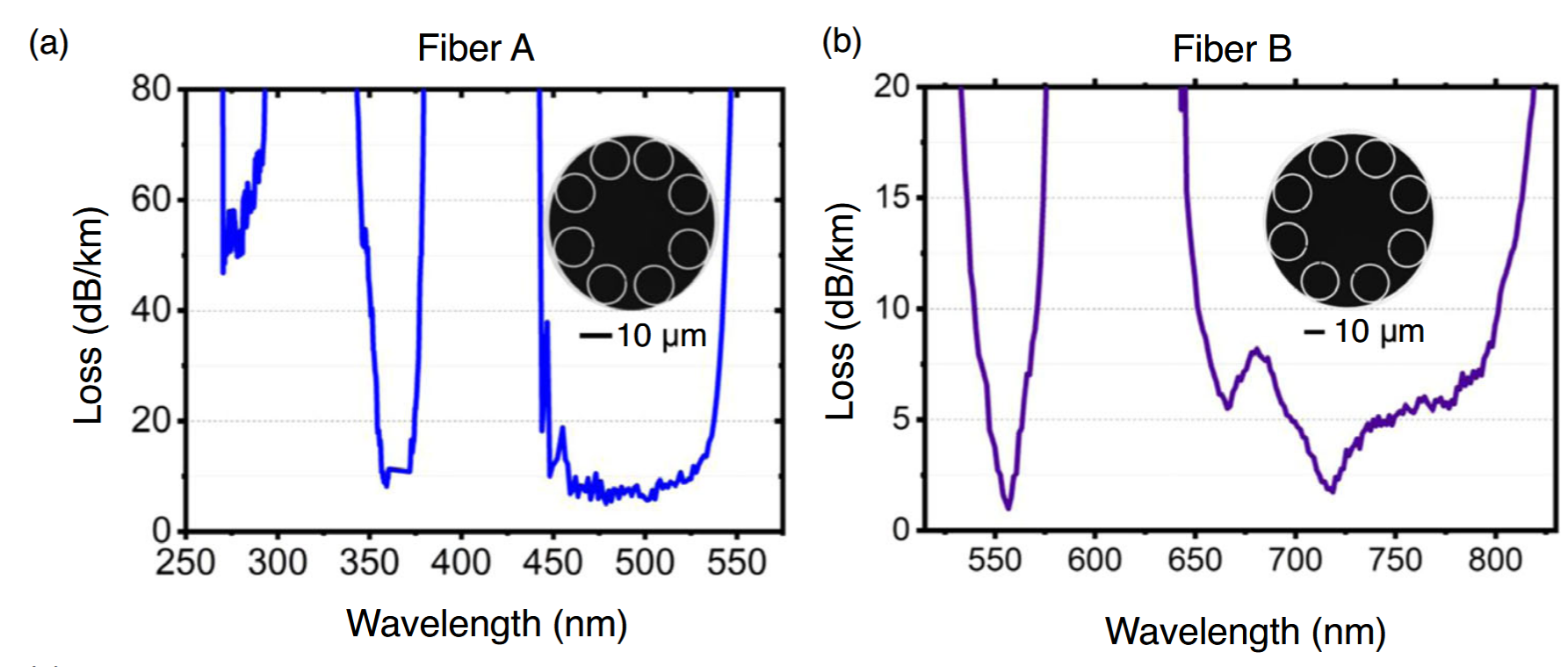}
        \label{fig:AR_upper}
    \end{subfigure}
    \hspace{0.015\textwidth} 
    \begin{subfigure}[b]{0.38\textwidth}
        \centering
        \includegraphics[width=\textwidth]{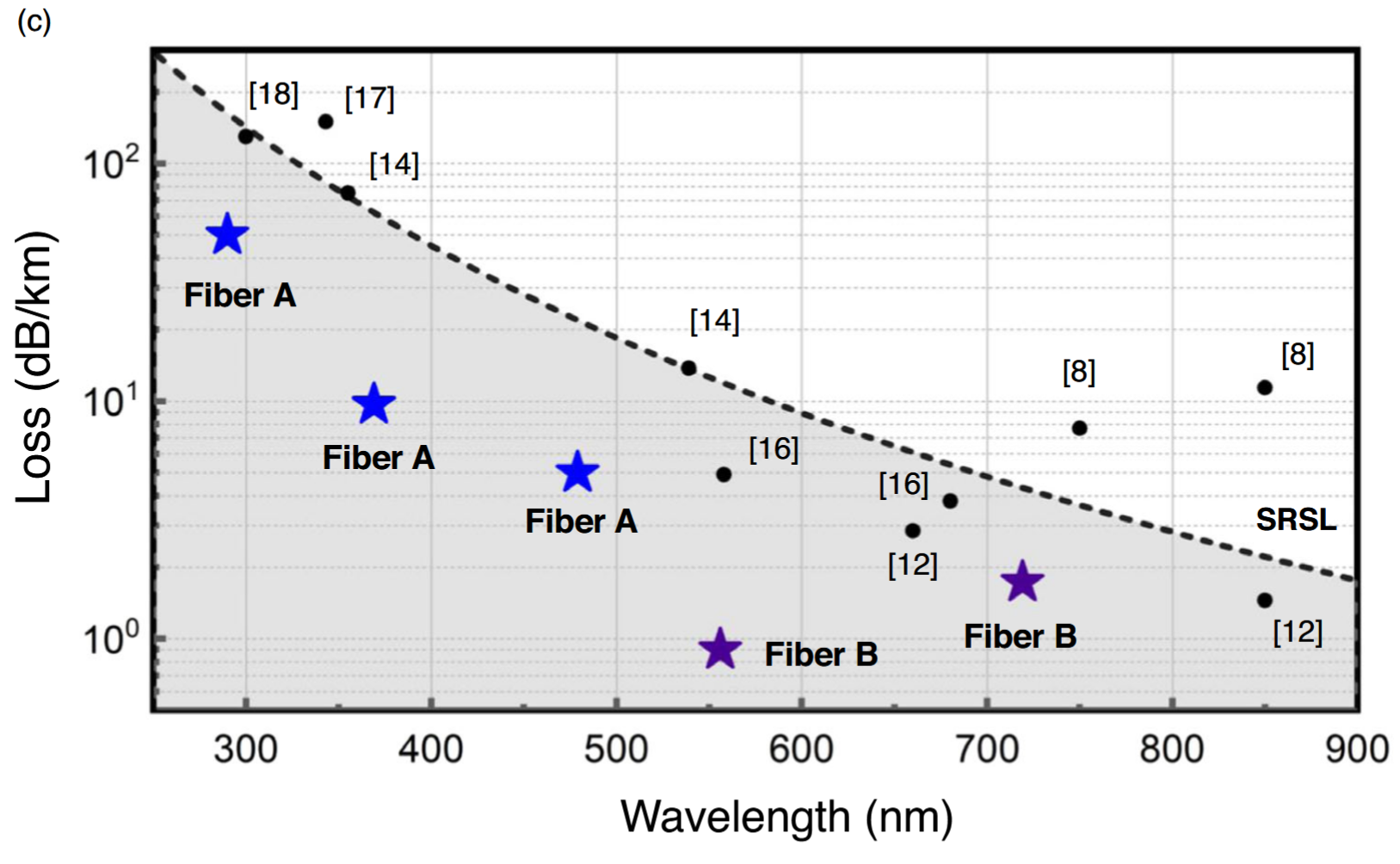}
        \label{fig:AR_lower}
    \end{subfigure}
    \caption{Graphs from reference \cite{osorio2023hollow} highlighting very low losses in already existing IC-HCFs by GloPhotonics.}
    \label{fig:existing_AR}
\end{figure}

While astronomy has traditionally relied on solid-core silica fibers, recent advances in hollow-core fiber fabrication have demonstrated ultra-low losses even in the short-wavelength visible regime. In particular, Osorio \textit{et al.} \cite{osorio2023hollow} report inhibited-coupling fibers with reduced core surface roughness, achieving a measured attenuation of approximately 9.7~dB/km at 369~nm (11$\%$ loss over 50~m). Figure \ref{fig:existing_AR} illustrates the measured attenuation across several fiber designs and wavelengths. By comparison, conventional FBP Polymicro fibers of the same 50~m length exhibit much higher losses: 360~nm: 51.14~dB/km (44.5$\%$ of loss), 375~nm: 36.76~dB/km (34.4$\%$ of loss) (see Table \ref{tab:fiber_throughput}).

In another publication \cite{sakr2020hollow}, nested anti-resonant nodeless fibers (NANFs) were reported to achieve attenuation of ~2.85~dB/km at 660~nm (3.23$\%$ of loss over 50 m of fiber length). By comparison, a conventional FBP Polymicro fiber exhibits a throughput loss of 9.98~dB/km (10$\%$ of loss over 50 m of fiber length) in the 600–700~nm range (see Table \ref{tab:fiber_throughput}). 

This demonstrates that already-existing hollow-core fibers can reduce losses by a factor of $\approx$4 in the visible regime, showing the capabilities of such fibers for enabling significantly higher throughput for astronomical applications.

%% file: a_Sec5.tex
\section{THROUGHPUT TESTING ON ROBOTIC POSITIONERS}
\subsection{Testing Methodology}
For the testing conducted in this section, FBP Polymicro fibers and GloPhotonics IC-HCFs were utilized. Figures \ref{fig:FBP_polymicro_spectrum} and \ref{fig:existing_AR} show the performance of the fibers in terms of loss as a function of the wavelength range that the fibers can support. The GloPhotonics fibers obtained are similar to the one represented as the fibers in Figure \ref{fig:existing_AR}. Due to their different light guiding mechanism, IC-HCFs operate in windows where they exhibit very low losses over a particular wavelength range. FBP Polymicro fibers support a much wider wavelength range, but suffer severely when it comes to shorter wavelengths, where IC-HCFs can play a fundamental role. 

The loss-to-wavelength characteristics presented in the previous sections give a very good idea about the performance of the different fibers, but it is essential to conduct the testing in the same operational conditions where these fibers will be installed. Therefore, in this section, we present the throughput testing of the fibers after having them installed on the prototype of 6 robotic positioners developed by the Swiss company MPS Micro Precision Systems AG. The robotic positioners are SCARA robots with two arms (alpha and beta) representing two degrees of freedom. 

The testing is conducted on two FBP Polymicro fibers of different dimensions. The first fiber has a core of 100 $\mu m$, a cladding of 120 $\mu m$, and a polyimide coating of 140 $\mu m$. The second fiber is a fiber acquired from the DESI project and has special dimensions with a core of 107 $\mu m$, a cladding of 140 $\mu m$, and a polyimide coating of 170 $\mu m$. Both Polymicro fibers have a numerical aperture (NA) of 0.22. The GloPhotonics fiber tested has an NA of about 0.01/0.02 with a core of 23 $\mu m$, a cladding of 180 $\mu m$, and an acrylate coating of 530 $\mu m$. 

A previous study \cite{huang2020bending} showed the performance of low-NA fibers in terms of bending and highlighted severe deterioration in performance when compared to high-NA fibers. Therefore, it was essential to conduct this testing on the positioners to understand the behavior of the fibers when subjected to bending or twisting. 

\begin{figure}[!t]
    \centering
    \begin{subfigure}[b]{0.33\textwidth}
        \centering
        \includegraphics[width=\textwidth]{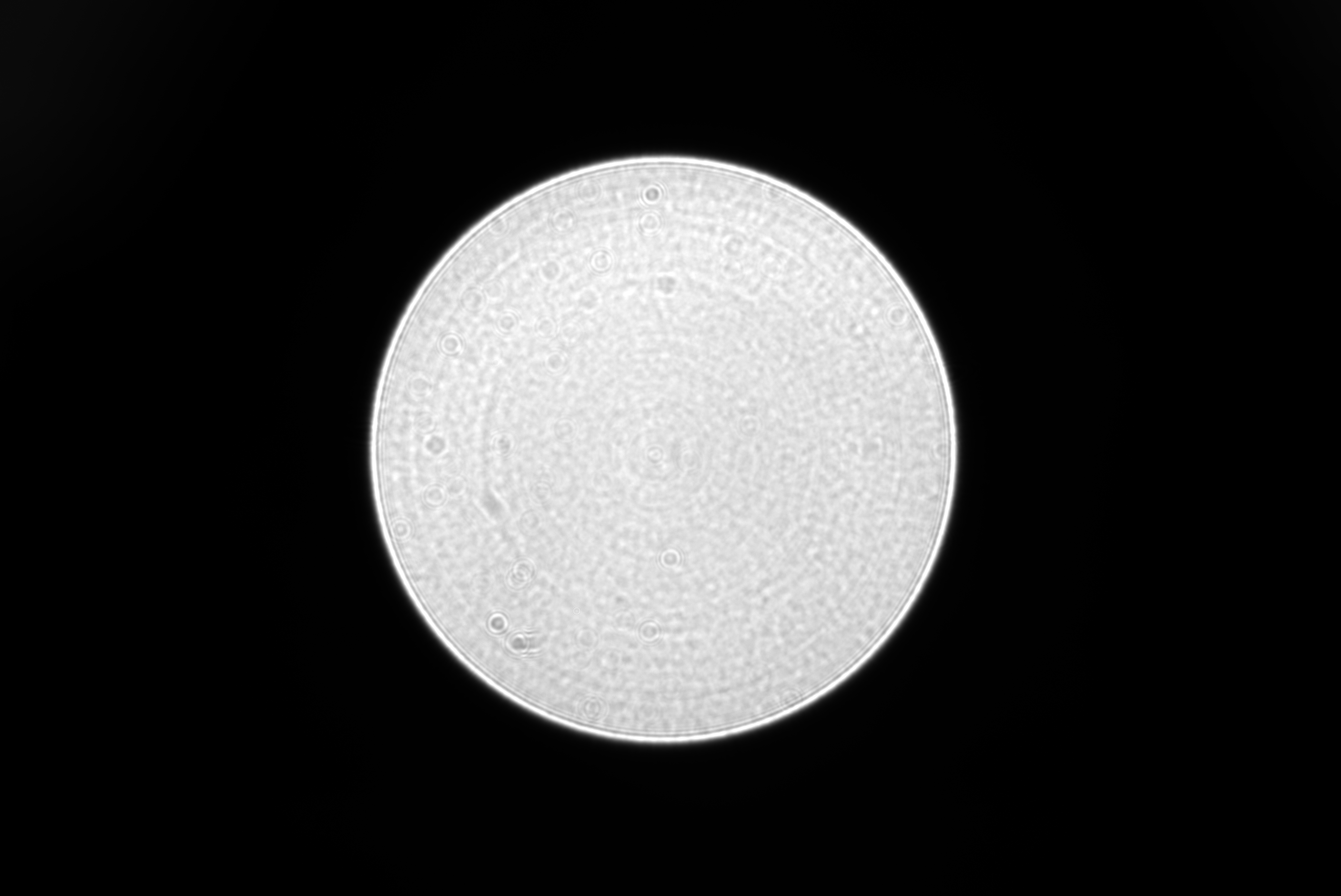}
        \label{fig:DESI_spot}
    \end{subfigure}
    \hspace{0.015\textwidth} 
    \begin{subfigure}[b]{0.33\textwidth}
        \centering
        \includegraphics[width=\textwidth]{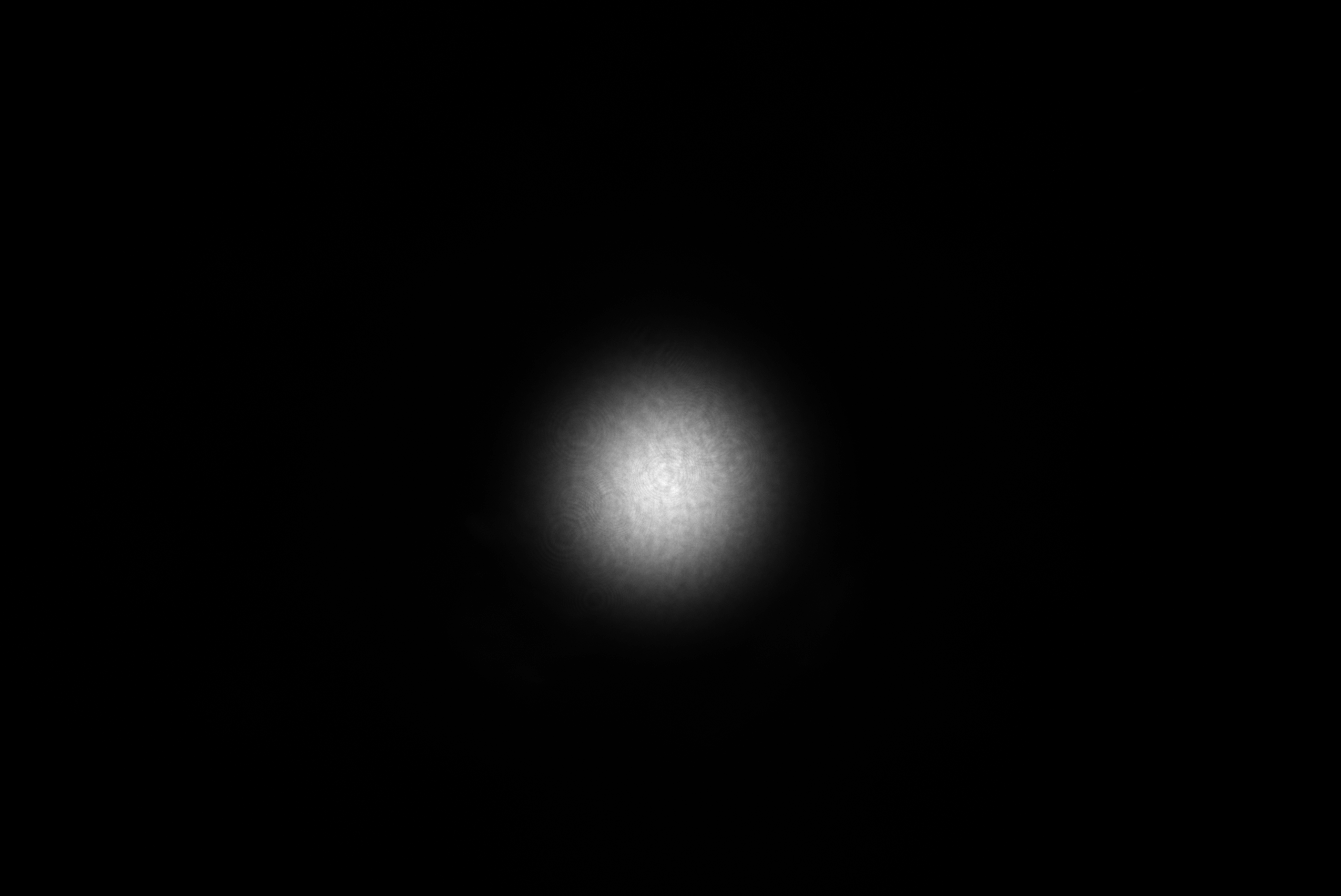}
        \label{fig:Glo_spot}
    \end{subfigure}
    \caption{Illuminated spots at the output of the optical fibers; (a) DESI Polymicro Fiber, (b) GloPhotonics Fiber.}
    \label{fig:lightspots}
\end{figure}

The test setup consists of an LED source at 625 nm that is fiber-coupled and directly connected to the optical fibers under test from the rear end. The light is then projected onto a Basler camera (Aca5472-17um) that has a Fujinon objective with a focal length of 35 mm attached to it. Due to the multi-modal nature of the fibers, a uniform illuminated spot is obtained. The different dimensions of the fibers result in different light spot sizes as shown in Figure \ref{fig:lightspots}. It should be noted that the experiment is highly sensitive to the uniformity of the light spot, as a non-uniform light spot will result in erroneous non-repeatable measurements.

The robotic positioners holding the optical fibers are commanded to make moves of 30 degrees for each arm going from 0 to 360 degrees in alpha and from 0 to 180 degrees in beta. For each alpha angle, beta completes its range, eventually leading to the acquisition of about 91 images. Each image is analyzed for each position of the robot, and a summation of the intensity of all pixels representing the encircled energy of the light spot is estimated as shown in Equation \ref{eq:ee_actual}. The throughput percentage is then calculated in percentage using Equation \ref{eq:throughput} as a function of the encircled energy at one position ($EE_{position}$) divided by the global maximum ($EE_{max}$) of all positions. The measurement is relative and no absolute throughput is calculated, as what is important is to infer the effect of the bending, twisting, or pinching on the fibers. 

\begin{equation}
    EE_{position} = \sum_{(x,y) \in \text{LightSpot}} I(x,y)
    \label{eq:ee_actual}
\end{equation}

\begin{equation}
    \text{Throughput (\%)} = \left( \frac{EE_{position}}{EE_{max}} \right) \times 100
    \label{eq:throughput}
\end{equation}

\subsection{Throughput Results}
\begin{figure}[!b]
    \centering
    \includegraphics[width=0.9\linewidth]{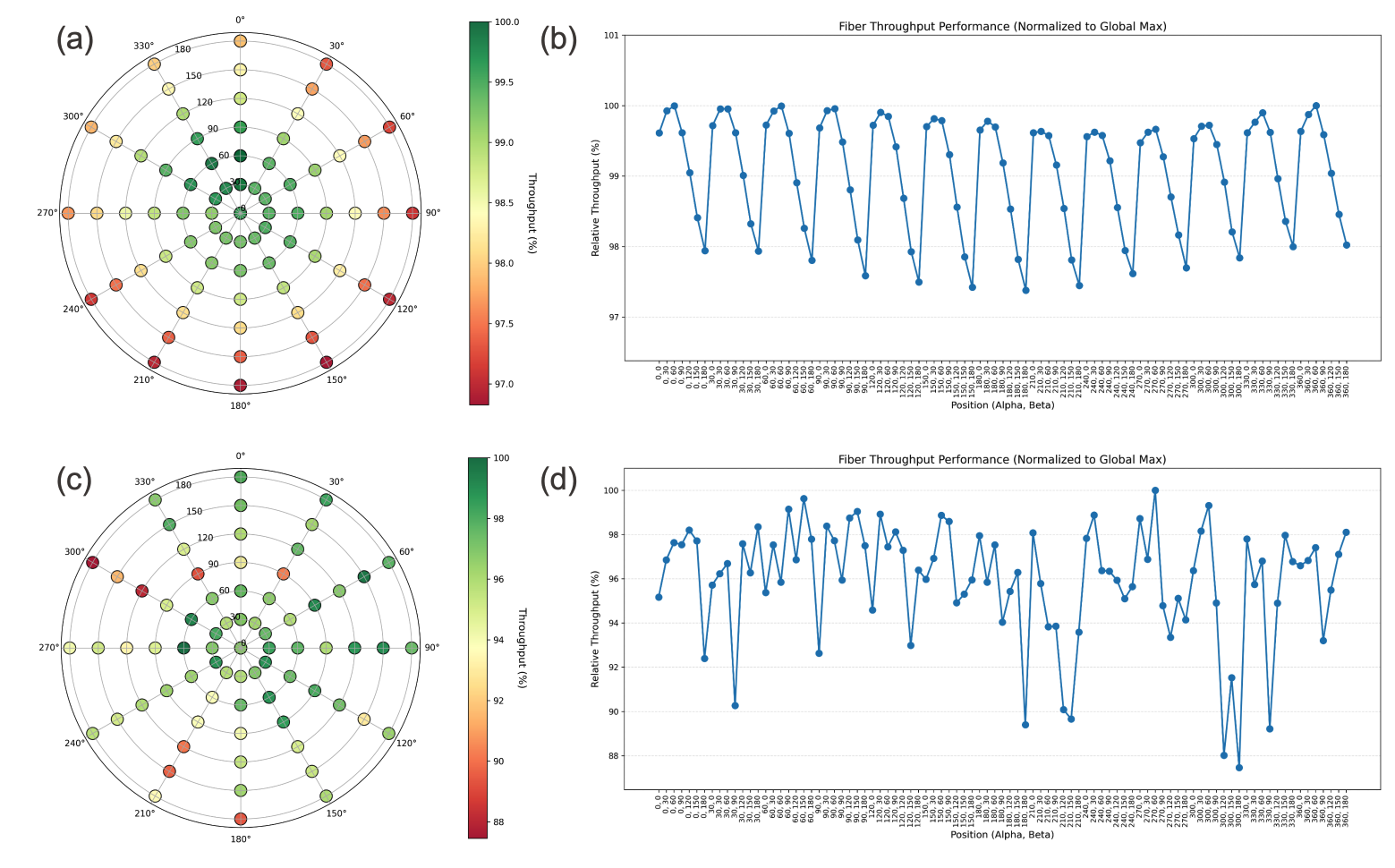}
    \caption{Graph showing throughput results for DESI Polymicro FBP fibers (a) and (b), and for GloPhotonics IC-HCF (c) and (d). The graphs indicate the throughput behavior as a function of the position in alpha and beta.}
    \label{fig:Throughput_Results}
\end{figure}

This section presents the throughput results obtained from the measurements conducted on the different fibers namely, the two FBP Polymicro fibers as well as the GloPhotonics IC-HCF. Table \ref{tab:fiber_throughput_results} shows the lowest throughput value reached during the movement of the robotic positioner over the various alpha and beta positions. This is further illustrated in Figure \ref{fig:Throughput_Results} focusing only on the results of the throughput measurements conducted for the FBP Polymicro DESI fibers and the GloPhotonics IC-HCF. The graphs show the evolution of the throughput as a function of the alpha and beta positions of the robotic arm. The values fluctuate, and as can be seen in Table \ref{tab:fiber_throughput_results}, the lowest value reached by FBP100 Polymicro fiber reaches 77\%, followed by the GloPhotonics IC-HCF having its lowest value at 86\%, and finally the FBP Polymicro used for the DESI instrument has its lowest value at 97\%. This measurement is relative to the global maximum reached by each fiber, so these results don't indicate absolute performance. They give an indication about how each fiber can withstand the movement of the robotic positioner's arm and maintain high performance.

It should be noted that the manner the fibers are screwed on the robotic positioners may affect the results by a few percents. For example, further testing was conducted on the DESI fiber where it was secured at a different position from the previous testing, and it showed an increment of the values by 2\%. This indicates that securing the fiber on the positioner needs to be done in such a manner that the fibers are relaxed to prevent them from being twisted. This, however, is a more mechanical aspect of the results, and therefore is out of the scope of this work. Since this paper focuses essentially on the quality of the fibers tested, we just aim to compare the performance of the fibers to each other under the same conditions.   

\begin{table}[h!]
    \centering
    \caption{Throughput of different fibers when tested on the robotic positioners.}
    \label{tab:fiber_throughput_results}
    \begin{tabular}{l *{9}{c}} 
        \toprule
        \toprule
        & \multicolumn{1}{c}{Lowest Throughput Value} \\
        \midrule
        FBP100 Polymicro & 77\% \\
        FBP Polymicro (DESI) & 97\% \\
        GloPhotonics IC-HCF & 86\% \\
        \bottomrule
        \bottomrule
    \end{tabular}
\end{table}

When comparing the FBP100 Polymicro fiber with the DESI Polymicro fiber, we believe that the superiority of the DESI fiber comes from its larger coating (170 $\mu$m compared to 140 $\mu$m), giving the fiber more stability to bend and twists that can occur while the positioner is moving. The GloPhotonics IC-HCF showed impressive stability in the bends and twists even though it has a smaller numerical aperture. The comparison presented in \cite{huang2020bending} shows that for fibers with 0.22 NA the throughput loss, depending on the input focal ratio and the bending radius, varies from 1$\%$ to 3$\%$. The throughput of the fiber with 0.12 NA varies from 20$\%$ to 30$\%$. Even though the GloPhotonics IC-HCF has a much lower NA (about 0.01/0.02), it was more resistant to twists and bends and performed better (due to its different light guiding mechanism) than the Polymicro fiber with 0.12 NA shown in the work by Huang et al. \cite{huang2020bending}

%% file: a_Sec6.tex
\section{DISCUSSION AND CONCLUSION}
This paper presents a comparison in throughput performance of FBP Polymicro Fibers and GloPhotonics Hollow-core Fibers. All fibers have been tested on a moving robotic positioner to evaluate the effect of bending, twisting, or pinching due to the movement of the positioner. The results show FBP100 exhibited the lowest performance in terms of the variation from the global maximum reaching as low as 77$\%$. It is then followed by the GloPhotonics HCF with the lowest throughput reaching 86$\%$. The DESI fiber was the sturdiest of the three with the lowest throughput at 97$\%$. 

We believe that the difference between FBP100 and the DESI fiber lies in the fact that the polyimide coating of the DESI fiber is thicker making the fiber more rigid. It is very interesting to see that the GloPhotonics HCF performed better than the FBP100 despite it having a lower NA and acrylate coating. Even though the difference between the DESI fiber and the GloPhotonics HCF is 11$\%$ (0.5 dB), the fact that the absolute throughput at shorter wavelengths is about 4x (6 dB) higher indicates the great potential of such hollow-core fibers to be investigated further for astronomical instruments focused on the short wavelength regime. 

Further experiments need to be conducted on the fibers such as focal ratio degradation (FRD) which is a vital parameter for etendue preservation in astronomical instruments. Due to its rather small core dimension, the GloPhotonics HCF is a few-mode fiber. To test the FRD of the optical fibers, there are two common test setups used, namely the collimated-beam ring test and the full-cone test. Attempts of FRD testing were conducted using a ring test, but due to the few-mode nature of the fiber, it was impossible to obtain a clear ring for the analysis. Accordingly, the measurements need to carried out using a full-cone setup or alternatively larger core dimensions of the hollow-core fiber could be also investigated. 

For a theoretical estimation, given the architectural differences between the GloPhotonics IC-HCF and the standard Polymicro FBP solid-core fiber, the FRD performance is expected to diverge significantly due to their distinct guiding mechanisms and material interactions. In a traditional solid-core step-index fiber like the Polymicro FBP, FRD is primarily driven by internal scattering (micro-bending), surface roughness at the core-cladding interface, and bulk inhomogeneities in the silica. As the input focal ratio ($f_{in}$) increases (narrower beam), the relative spreading caused by these imperfections becomes more pronounced, leading to a significant degradation of the output focal ratio ($f_{out}$). By contrast, the GloPhotonics IC-HCF is expected to exhibit superior etendue preservation because the optical power is confined within a hollow core (air or vacuum), and thus the light avoids the bulk scattering and refractive index fluctuations inherent in solid silica. This fundamentally limits the "diffusion" of light into higher-order modes during propagation. Additionally, the Inhibited Coupling mechanism typically shows lower sensitivity to small-scale mechanical perturbations compared to standard total internal reflection fibers. In solid-core fibers, external stresses cause mode coupling that broadens the output cone; in IC-HCF, the modal field is more robustly isolated from the cladding structure.The IC-HCF also has fewer available modes to couple into, and accordingly, the light is "forced" to remain in its launched state. In the large-core Polymicro fiber, there is a dense continuum of modes, allowing even tiny perturbations to spread the energy across a wider angular distribution.

Another important remark about IC-HCFs is that their architecture inherently necessitates a smaller core diameter, typically around several 10s of $\mu$m, compared to the 100$\mu$m standard found in solid-core fibers like the FBP100. This smaller aperture is a physical consequence of the Inhibited Coupling mechanism and the need to maintain modal purity. For astronomical applications, this introduces a significant challenge: the requirement for highly precise alignment and injection. While it is not immediately intuitive to transition from large-core fibers to such restricted dimensions, modern Adaptive Optics (AO) systems make this feasible. By correcting atmospheric turbulence to produce a diffraction-limited PSF, an AO system can concentrate stellar light into a spot size significantly smaller than the IC-HCF core. Thus, while the HCF demands a more sophisticated front-end setup, the payoff is a fiber link that is far less susceptible to the modal noise and focal ratio degradation that plague larger, solid-core counterparts.

In conclusion, hollow-core fibers are already proving their worth across a diverse range of high-precision industries, from telecommunications to laser power delivery. Our results indicate that they are now poised to present the next major breakthrough in astronomical instrumentation, particularly for observations in the short-wavelength (UV/Blue) regime. Despite current limitations regarding core size and the need for specialized injection systems, the IC-HCF demonstrated remarkable resilience when integrated into the dynamic environment of a moving robotic positioner. By offering up to 4x higher absolute throughput at short wavelengths and exhibiting better relative stability than standard solid-core fibers like the FBP100, these fibers provide a promising solution for the next generation of high-resolution spectrographs and fiber-fed survey instruments.